\documentclass[doublecol]{epl2} 
\usepackage{amsbsy}
\usepackage{amsmath}

\def\mbb{b\bar{b}}
\def\bb{$\mbb$}
\def\muns{\Upsilon(nS)}
\def\mus{\Upsilon(1S)}
\def\muss{\Upsilon(2S)}
\def\musss{\Upsilon(3S)}
\def\uns{$\muns$}
\def\us{$\mus$}
\def\uss{$\muss$}
\def\usss{$\musss$}

\def\mcp{\chi_b(1P)}
\def\mcpp{\chi_b(2P)}
\def\mcppp{\chi_b(3P)}

\def\cp{$\mcp$}
\def\cpp{$\mcpp$}
\def\cppp{$\mcppp$}
\def\mRprel{R^\text{prel}_{AA}}
\def\Rprel{$\mRprel$}

\title{The influence of the $\boldsymbol \chi_{\boldsymbol b}$(3P) state
on the decay cascade of \\ bottomium 
in PbPb collisions at LHC energies}
\shorttitle{The influence of the $\chi_b(3P)$ state} 

\author{F.  Vaccaro \and F. Nendzig \and G. Wolschin}
\shortauthor{F. Vaccaro\etal}

\institute{                    
  Institut f{\"ur} Theoretische
Physik
der Universit{\"a}t Heidelberg, Philosophenweg 16, D-69120 Heidelberg, Germany, EU\\
  
}

\abstract{We investigate the decay cascade of the $\Upsilon(nS)$ meson including the newly found $\chi_b(3P)$ state in $pp$ and PbPb collisions at LHC energies. The main goal is to quantitatively 
determine the
additional suppression of the $\Upsilon(nS)$ states in PbPb collisions relative to $pp$ at
LHC energies of $\sqrt{s_{NN}}=2.76$ TeV when the $\chi_b (3P)$ state is included together with the 
$\Upsilon(nS)$ and $\chi_b(1P, 2P)$ states. It is found that the suppression of $\Upsilon(1S)$ in PbPb collisions relative to $pp$ is increased by at most 7 {\%} through the inclusion of $\chi_b(3P)$,
whereas the suppression factors for the $\Upsilon(2S)$ and $\Upsilon(3S)$ states do not change significantly.}

\begin{document}

\maketitle

\section{Introduction}
Heavy quarkonia are of great current interest as a probe for the detailed investigation of the quark-gluon plasma. Such a novel state of matter is believed to be formed in relativistic heavy-ion collisions at RHIC and LHC energies. Particularly well suited for these investigations is the bottomium system due to the large $b$--quark mass, the extraordinary stability of its ground state in the quark-gluon plasma environment, and a negligible recombination contribution.

Its suppression in PbPb collisions at the current LHC energy of 2.76 TeV per nucleon pair as compared to $pp$ collisions at the same energy has recently been measured with high precision by the CMS collaboration
\cite{CMS2011a,CMS-2012} through the decay of the $\Upsilon(nS)$ states into dimuon pairs. It was found that the $\Upsilon(1S)$ ground state suppression factor is 56\% in minimum-bias collisions, and that the suppression increases monotonically with centrality. The suppression of the excited states is significantly larger: Only 12\% survive in minimum bias for the $\Upsilon(2S)$ state, 3\% for the $3S$ state.

Theoretical considerations reveal accordingly considerable suppression of the $\Upsilon(2S, 3S)$ $l=0$ excited states and also of the $\chi_b(1P, 2P)$ $l=1$ states through screening, gluon-induced dissociation and collisional damping \cite{ngw13}, although a discrepancy to the large experimental suppression remains presently unexplained. In contrast, the $\Upsilon(1S)$ ground state is very stable with respect to screening of the real part of the quark-antiquark potential \cite{ms86}, but it is affected by gluodissociation \cite{brezinski-wolschin-2012}, and collisional damping (imaginary part of the potential \cite{str12}). 

In addition, the reduced occupation of the excited states modifies the feed-down of these states to the $1S$ ground state that occurs subsequently to the interactions within the fireball. As a result of the calculated thermal suppression in the fireball, plus the modification of the feed-down cascade, good agreement between the experimental CMS result for the \us\ state suppression \cite{CMS-2012} and our theoretical model \cite{ngw13} has been achieved.

In the calculations for the effect of the feed-down cascade presented in \cite{ngw13}, we had considered the five $\Upsilon(nS)$ and $\chi_b(nP)$ states mentioned above. We had neglected spin-flip transitions to the $\eta_b$ and $h_b$ states since the transition rates are negligible. The purpose of the present work is to investigate the additional effect of the $\chi_b (3P)$ state on the cascade, and in particular, on the possible modification of the ground state suppression.

The $\chi_b (3P)$ state was recently observed for the first time by the ATLAS Collaboration in 7 TeV $pp$ collisions at the LHC \cite{atlas-2013}. The discovery was subsequently confirmed by the D0 Collaboration based on earlier Tevatron $p\bar{p}$ data at 1.96 TeV \cite{aba12}. 

There are, however, no measured branching ratios for the radiative and hadronic decays to the low-lying $\Upsilon$ and $\chi_b$ states available. Hence we rely on the theoretical work \cite{dasi87} for the radiative decay widths in order to calculate the effect of the $\chi_b (3P)$ state on the decay cascade, and thus on the suppression of the $\Upsilon(1S)$ ground state in PbPb collisions at LHC energies.

The calculation of the feed-down cascade is considered in the next section. We subsequently describe how to include the effect of the $\chi_b (3P)$ state into the cascade calculation. 
Then we investigate its influence on the $\Upsilon(1S)$ ground state suppression factor in minimum-bias and centrality dependent PbPb collisions at LHC energies of 2.76 TeV. Here we take the preliminary suppression factors that are due to screening, damping, and gluodissociation from our previous work \cite{ngw13}, but add the effect of the feed-down cascade including the $\chi_b (3P)$ state. The conclusions regarding the additional suppression through consideration of the $\chi_b (3P)$ state are drawn at the end.

\section{The feed-down cascade}
\begin{figure}[tph]
\begin{center}
\includegraphics[width=8.4cm]{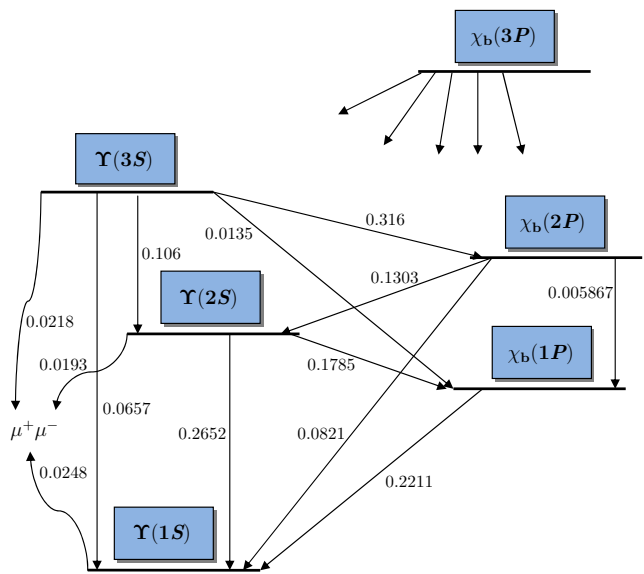}
\caption{\label{fig1}(Color online) Branching ratios for decays within the bottomium family 
and into $\mu^\pm$-pairs according to PDG 2012 \cite{pdg2012}. Branching ratios for the newly found $\chi_b(3P)$ state are still unknown; partial $\gamma$ decay widths are taken from theory \cite{dasi87} as discussed in the text.} 
\end{center}
\end{figure}

We consider the effect of the feed-down cascade in $\Upsilon$ suppression in two distinct but intimately related physical situations. First, in $pp$ collisions the final population of the \uns\ states is measured through \text{dimuon} decays \cite{CMS2011a}. To obtain the initial populations, an inverted decay cascade calculation is performed that includes all possible decay paths to reach the measured final states, see Fig.\ref{fig1}.
These initial populations are then used as an input for PbPb at the same c.m. energy. Hence they are the starting point to describe the thermal suppression in the hot medium.

The thermal suppression due to screening, damping and gluodissociation in the quark-gluon plasma (QGP) generated in PbPb collisions at the same c.m. energy is considered in \cite{ngw13} for five bottomium states, and in this work for six bottomium states. We obtain preliminary suppression factors \text{\Rprel} for these six states in PbPb, and
 subsequently calculate the decay cascade into lower-lying bottomium states and into $\mu^+\mu^-$ pairs that are used for the detection of the states in order to obtain final suppression factors $R_{AA}$, which may be compared directly to the available data. It will then be possible to deduce the effect of the $\chi_b(3P)$ state on the cascade, and on the overall suppression of the $\Upsilon(nS)$ states.
 
Fig. \ref{fig1} displays the decays within the bottomium family and into dimuon pairs with experimental branching ratios taken from the 2012 PDG listings (which differ in some details from previous listings) \cite{pdg2012}.
For the $\chi_b(3P)$ state, the partial radiative decay widths into the $\Upsilon(nS)$ states as taken from theory \cite{dasi87} are shown in table \ref{tab.1} since experimental values are not yet available. The unknown hadronic decay widths into the lower $l = 1$ states are assumed to be negligible. 

The separation of 
 processes inside the fireball from the subsequent decay cascade is justified by the very
different time scales involved:  The fireball cools within about $\simeq$ 10 fm/$c$, while the decay cascade occurs on time scales $\sim 10^{4}$ fm/$c$ for electromagnetic as well as for hadronic decays since the corresponding partial decay widths are in the 10 keV range. As an example, the hadronic decay \uss\ to \us\ has a partial width of 8.6 keV, it occurs on a time scale of $2.3\cdot 10^4$ fm/$c$. Thermal decay widths of the same states in the hot thermal medium due to gluodissociation and damping are in the MeV range, see \cite{ngw13} for comparison.

Before calculating explicitly the effect of the $\chi_b(3P)$ state, we reconsider the cascade calculation as performed in \cite{ngw13} for five states.
We denote \bb\ states by $I=(nl)$, averaging over hyperfine states for $l=1$. The branching ratio for the decay of state $J$ 
into state $I$ including all indirect decays with intermediate \bb\ states is $(\mathcal{C}_{IJ})$ ($I \leq J$) . The initial and final populations $N^i(I)$ and $N^f(I)$ of state $I$ in $pp$ and PbPb collisions are then connected by
\begin{align}
 N^f_{pp}(I) = \sum_{I \leq J} \mathcal{C}_{IJ} N^i(J),	\\
 N^f_\text{PbPb}(I) = \sum_{I \leq J} \mathcal{C}_{IJ} N^i(J) \mRprel(J).
\end{align}
Inverting the first equation gives us the initial populations $N^i(J)$ of the \uns\ states in $pp$ collisions.
Since bottomia states are believed to be created in the early stages of heavy-ion collisions through the same hard collision processes as in $pp$, we take these initial populations (scaled with the number of binary collisions) for PbPb collisions at the same c.m. energy.

We define the number of \uns\ states that decay into dimuon pairs
\begin{equation}
 N^f_{\mu^\pm}(nS) = \mathcal{B}(nS\rightarrow \mu^\pm) N^f_\text{PbPb}(nS),
\end{equation}
with the corresponding branching ratio $\mathcal{B}(nS\rightarrow \mu^\pm)$.

We take $N^f_{\mu^\pm}(nS)$ from the
2011 CMS $pp$ data \cite{CMS-2012} and consider that 27.1\% and 10.5\% of the \us\ population come from \cp\ and \cpp\ decays,
respectively \cite{aff00}. These CDF results are, however, obtained from $p\bar{p}$ collisions at 1.8 TeV
with a transverse momentum cut $p_T^{\Upsilon}>8.0$ GeV/$c$. Hence
it would be desirable to confirm the \us\ populations from $\chi_b$ decays in new $pp$ measurements at 2.76 TeV, which are not yet available.

\begin{figure}[t]
	\centering
\includegraphics[width=8.4cm]{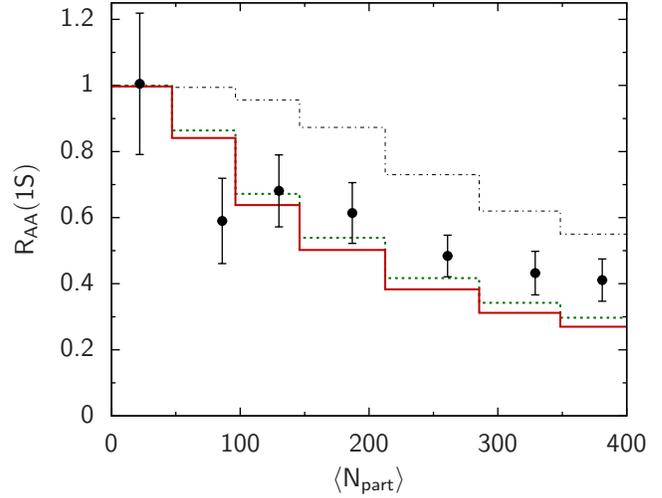}
\caption{\label{fig2}(Color online) Suppression factor $R_{AA}$ for the \us\ ground state calculated for $2.76$ TeV PbPb collisions in centrality bins 50--100\%, 40--50\%, 30--40\%, 20--30\%, 10--20\%, 5--10\%, 0-5\% and compared with CMS data \cite{CMS-2012}. The upper dash-dotted line 
is the preliminary suppression factor \Rprel(1S) ($t_\text{QGP} = 6$ fm/$c$) calculated 
from collisional damping, gluodissociation and screening in \cite{ngw13}. 
The dotted curve includes the feed-down cascade with four excited states as in \cite{ngw13}. The solid curve considers in addition the effect of the $\chi_b(3P)$ state, which causes a slight additional suppression.}
\end{figure}

 The initial populations are then obtained in units of $\mathcal{B}(nS\rightarrow
\mu^\pm) N^f_{pp}(1S)$ as $N^i(1S) = 16.2$, $N^i(1P) = 43.7$, $N^i(2S) = 20.3$, $N^i(2P) = 45.6$, $N^i(3S) = 18.8$. The
final suppression factor is calculated as $R_{AA}(nS) = N^f_{\mu^\pm}(nS)/N^f_{pp}(nS)$ or
\begin{equation}
 R_{AA}(nS) = \mathcal{B}(nS\rightarrow \mu^\pm) \frac{\sum_{nS \leq J} \mathcal{C}_{IJ} N^i(J) \mRprel(J)}{\sum_{nS
\leq J} \mathcal{C}_{IJ} N^i(J)}.	
\label{finalRAA}
\end{equation}
We had calculated the preliminary suppression factors \text{$\mRprel(J)$}  that enter the above calculation in \cite{ngw13} both for minimum bias, and as a function of centrality. The latter are included in Fig. \ref{fig2} as the upper dash-dotted curve for the \us\ state. Due to the reduced occupation of the excited states in the QGP environment of a PbPb collision at LHC energies relative to the scaled population in a $pp$ collision at the same energy, the suppression is substantially larger as a consequence of reduced feed-down. 

In a comparison of the calculated \us\ suppression that includes feed-down from \uns, \cp\  and \cpp\ with the CMS data \cite{CMS-2012}, we found reasonable agreement  for a \bb\ formation time of 0.1 fm/$c$, and a QGP lifetime of 6-8 fm/$c$. Fig. 2 shows the results for 6 fm/$c$ as dotted curve.

\section{Inclusion of the $\boldsymbol{\chi_b(3P)}$ state}
To consider the effect of the $\chi_b(3P)$ state on the suppression of the \uns\ states, we have to know the branching ratios into the lower \bb\ states - in particular, for the electromagnetic $E1$ transitions to the \uns\ states. 
\begin{table}
\caption{Partial decay width of the $\chi_{bj}(3P)$ hyperfine states in keV as evaluated in \cite{dasi87}.}
\label{tab.1}
\begin{center}
\begin{tabular}{llcr}
\hline

Decay mode &$\chi_{b0}(3P)$&$\chi_{b1}(3P)$&$\chi_{b2}(3P)$\\
\hline

$\gamma\Upsilon(1S)$ & 0.33 &1.7&4.6\\
$\gamma\Upsilon(2S)$ & 1.0 &2.4&4.4\\
$\gamma\Upsilon(3S)$ & 7.8 &9.4&11.4\\
\hline\\
\end{tabular}
\end{center}
\end{table}
Experimentally none of these ratios is known for this newly discovered state, but theoretical values for the partial decay widths of the radiative transitions have been calculated by Daghighian and Silverman \cite{dasi87} in a QCD-motivated relativistic formulation of radiative decays for quark-antiquark bound states. Since their results are in good agreement with experiment for cases where data are available,
it can be expected that the values as listed in Table \ref{tab.1} are reliable, and we use them in our subsequent calculation.

In order to simplify the cascade calculation, we average over the decaying $\chi_{bj}(3P)$ hyperfine states weighted with $(2j+1)$ that are listed in the table, to obtain partial decay widths of 3.159 keV into \us, 3.356 keV into \uss, and 10.333 keV into \usss, in total  16.848 keV for these three decays. It should be noted, however, that the total decay width of the \cppp\ state is unknown. Since the total width cancels out in the nuclear suppression factor $R_{AA}$ this does not represent a problem. 

 

We also assume -- as was proposed in \cite{aff00} -- that all $\Upsilon(3S)$ mesons come
from $\chi_{b}(3P)$ decays. With the theoretical radiative decay widths of \cite{dasi87}, this gives an upper limit of 6\%  \cite{aff00} for the fraction of $\Upsilon(1S)$ arising from $\chi_{b}(3P)$ decays, which we use as a boundary condition in our calculation. 

From the inclusion of the $\chi_{b}(3P)$ state we expect a more pronounced suppression of the \uns\ states: As the
relative population of the highly excited $3P$ state increases, fewer \bb\ particles survive the quark-gluon plasma and hence, the dimuon yield decreases -- in particular, from the \us\ state.
We now determine this additional suppression quantitatively.

\section{Effect of the $\boldsymbol {\chi_{b}(3P)}$ state on the suppression}
Using the theoretical partial decay widths from \cite{dasi87} and 6\% for the fraction of $\Upsilon(1S)$ arising from $\chi_{b}(3P)$ decays, the minimum bias suppression factor of the \us\ state is found to be $R_{AA} = 0.425$ for a Upsilon formation time of $t_\text{F}$ = 0.1 fm/$c$, and a QGP lifetime of $t_\text{QGP}$ = 6 fm/$c$.

The corresponding value for the ground state minimum-bias suppression factor that we calculated without consideration of the $\chi_{b}(3P)$ state is $R_{AA} = 0.455$ \cite{ngw13}; the CMS experimental result is $0.56\pm 0.08\pm 0.07$ \cite{CMS-2012}. 

Hence our result for the ground-state suppression factor is lowered by about 6.7\% when including the $\chi_{b}(3P)$ state, while the \uss\ minimum-bias suppression factor is reduced by only 1.4\%, and the reduction in the \usss\ suppression factor is even less. 

In all cases, there is thus more suppression when the $\chi_{b}(3P)$ state is included. Its
 effect is smaller for the higher excited states because the populations in the fireball differ only from the scaled $pp$ case if a QGP is present.
The higher excited states, however, are populated almost exclusively only in peripheral collisions where no quark-gluon plasma is formed and hence, the effect of the $\chi_{b}(3P)$ cancels out.
In more central collisions, color screening prevents their formation as discussed in \cite{ngw13}.

The influence of the $3P$ state on the centrality-dependent $1S$ suppression factor is seen in Fig.\ref{fig2}. It is negligible for 50-100\% peripheral collisions where thermal mechanisms in the fireball above $T_c$ are unimportant, and the effect of the $\chi_{b}(3P)$ state on 
$R_{AA}(1S)$ cancels out. It becomes gradually more important with increasing centrality, but then decreases again slightly for very central collisions where the $\chi_{b}(3P)$ state vanishes almost completely due to color screening.



We conclude that whereas the consideration of the newly detected $\chi_{b}(3P)$ state with respect to the suppression of Upsilon mesons in the quark-gluon plasma that is formed in heavy-ion collisions at LHC energies constitutes an interesting physical problem, its effect remains below the 7\% level for the \us\ state in minimum-bias collisions. Still, it  should be considered in future more detailed descriptions of bottomia suppression in the QGP.


\acknowledgments
This work has been supported by the IMPRS-PTFS, Heidelberg, and the ExtreMe Matter Institute EMMI.

\bibliographystyle{eplbib}
\bibliography{references}

\end{document}